\documentclass[twoside,preprintnumbers,amsmath,amssymb,pacs,showpacs,nofootinbib]{revtex4}
\usepackage{amsmath,amssymb,url}
\usepackage{graphicx,subfigure}
\usepackage[euler-hat-accent,euler-digits]{eulervm}
\usepackage{braket,euscript,enumitem}
\usepackage{fancyvrb}

\usepackage{xcolor}

\newcommand{\MSbar}{\overline{\mbox{MS}}}

\newcommand{\p}{\partial}

\newcommand{\lms}{\Lambda_{\overline{\mbox{\tiny{MS}}}}}
\begin{document}

\title{{\bf Accessing the topological susceptibility via the Gribov horizon}}
\author{D.~Dudal$^{\dag,\ddag}$, C.~P.~Felix$^{\dag}$, M.~S.~Guimaraes$^{\S}$, S.~P.~Sorella$^{\S}$  }
\email{david.dudal@kuleuven.be,caroline.felix@kuleuven.be, msguimaraes@uerj.br, silvio.sorella@gmail.com  }

\affiliation{$\dag$ KU Leuven Campus Kortrijk -- Kulak, Department of Physics, Etienne Sabbelaan 53 bus 7657, 8500 Kortrijk, Belgium\\$^\ddag$ Ghent University, Department of Physics and Astronomy, Krijgslaan 281-S9, 9000 Gent, Belgium\\ $^\S$ Instituto de F\'isica Te\'orica, Rua S\~ao Francisco Xavier 524, 20550-013, Maracan\~a, Rio de Janeiro, Brasil}

\pacs{}
\begin{abstract}
The topological susceptibility, $\chi^4$, following the work of Witten and Veneziano, plays a key role in identifying the relative magnitude of the $\eta^{\prime}$ mass, the so-called $U(1)_{A}$  problem. A nonzero $\chi^4$ is caused by the Veneziano ghost, the occurrence of an unphysical massless pole in the correlation function of the topological current $K_{\mu}$. In a recent paper, \cite{Kharzeev:2015xsa}, an explicit relationship between this Veneziano ghost and color confinement was proposed, by connecting the dynamics of the Veneziano ghost, and thus the topological susceptibility, with Gribov copies. However, the analysis of \cite{Kharzeev:2015xsa} is incompatible with BRST symmetry \cite{Dudal:2015khv}. In this paper, we investigate the topological susceptibility, $\chi^4$, in $SU(3)$ and $SU(2)$ Euclidean Yang-Mills theory using an appropriate Pad\'{e} approximation tool and a non-perturbative gluon propagator, within a BRST invariant framework and by taking into account Gribov copies in a general linear covariant gauge.

\end{abstract}
\maketitle

\section{Introductory remarks}
There are two important properties of QCD that are decisive in determining its particle spectrum: {\it confinement}, the fact that quarks and gluons are not observable as free particles, and {\it chiral symmetry breaking}, answering the question why  hadrons composed of $u$, $d$ or $s$ quarks or antiquarks are so massive while these (light) quark flavours themselves are almost massless.

Already 40 years ago, Gribov \cite{Gribov:1977wm} showed that the Faddeev-Popov construction is not valid at the non-perturbative level, i.e.~in the low energy limit where the coupling constant, $g$, is large. In this regime, we have Gribov copies, caused by  multiple intersections of gauge orbits with the  hypersurface corresponding to a given   gauge condition $f(A)=0$. This means that we have to deal in some way with equivalent field configurations obeying the same gauge fixing condition. Thus, in a non-perturbative non-Abelian gauge theory setting, the Faddeev-Popov procedure is incomplete as it stands.
For reviews, see \cite{Vandersickel:2012tz,Sobreiro:2005ec}. The Gribov problem for covariant gauges was also put on a mathematical footing in \cite{Singer:1978dk}, at the same time showing  that it cannot be avoided. Loosely speaking, Gribov copies imply that:
\begin{itemize}
\item
we are overcounting equivalent gauge configurations, since we have more than one gauge fixed configuration for each gauge orbit. This implies the Faddeev-Popov $\delta$-function implementing the gauge condition will have multiple zeros of its argument, complicating its interpretation as a unity being inserted into the a priori gauge invariant partition function.

\item
the Faddeev-Popov measure is ill-defined,  since there are zero-modes of the Faddeev-Popov operator when considering the infinitesimal copies. This implies a vanishing Faddeev-Popov (Jacobian) determinant. Indeed, considering 2 infinitesimally connected gauge configurations,\begin{equation}
\Tilde{A}^{a}_{\mu}=A^{a}_{\mu}+D^{ab}_{\mu}\theta^{b},
\label{gauge_equiv}
\end{equation}
$\Tilde{A}^{a}_{\mu}$ will obey the same (linear) gauge condition, $\p_\mu A_\mu^a=f^a$, as $A_\mu^a$ if
\begin{equation}
-\partial_{\mu}D^{ab}_{\mu}\theta^{b}=0,
\label{gange_A}
\end{equation}
i.e.~whenever the Faddeev-Popov operator exhibits normalizable zero modes, see \cite{Sobreiro:2005ec} for some explicit example.

\end{itemize}

To solve this problem, Gribov proposed  to restrict the domain of integration in the path integral to a certain region $\Omega$ in field space, called the Gribov region,  which is  free from infinitesimal Gribov copies:
\begin{equation}
\Omega=\{A_{\mu}^a;\  \partial_\mu A_{\mu}^a=0,\qquad \mathcal{M}^{ab}(A)=-\partial_\mu D_\mu^{ab}(A)>0 \}.
\label{GR}
\end{equation}
The Gribov region is the set of all field configurations which obey the Landau gauge, $\partial_\mu A_{\mu}^a=0$, and where the Hermitian Faddeev-Popov operator, $\mathcal{M}^{ab}(A)$, is positive. Inside the Gribov region, there are no infinitesimal copies, since $\mathcal{M}^{ab}(A)>0$.  The region $\Omega$  is also known to
be convex, bounded and intersected  at least once by every gauge orbit \cite{Dell'Antonio:1989jn}.  Its boundary, $\partial\Omega$, where the first vanishing  eigenvalue of  $\mathcal{M}^{ab}(A)$  (i.e.~the first zero-mode of Faddeev-Popov operator) appears, is called the first Gribov horizon.  Unfortunately, even if the restriction of the domain of integration in the functional integral to the region $\Omega$ allows to get rid of a large number of copies, there still remain additional copies inside $\Omega$ \cite{vanBaal:1991zw,semenov}. A subregion of $\Omega$, known as the fundamental modular region $\Lambda$, has been proven to be fully free from Gribov copies.  Though, unlike the Gribov region $\Omega$, a concrete setup to implement the restriction of the domain of integration in the functional integral to $\Lambda$ within a local and renormalizable framework is still far beyond our present capabilities. Therefore, we shall focus on the Gribov region $\Omega$, which enables us to already capture quite useful non-perturbative aspects, see \cite{applications} for recent applications.  Let us proceed by giving a sketchy overview of the construction of the local action which emerges from the restriction to $\Omega$, referring to \cite{Gribov:1977wm,Zwanziger:1989mf,Zwanziger:1992qr,Maggiore:1993wq,Dudal:2010fq,Capri:2015mna,Dudal:2007cw,Dudal:2008sp,Dudal:2011gd} for the specific details.

 In the Landau gauge, the effective implementation of the restriction to $\Omega$  is accomplished by means of the Gribov-Zwanziger action \cite{Gribov:1977wm,Zwanziger:1989mf,Zwanziger:1992qr},
\begin{eqnarray}
S&=&S_{YM}+S_{GF}+S_{GZ}.
\label{S_gamma}
\end{eqnarray}
$S_{YM}$ is the Yang-Mills action,
 \begin{equation}
S_{YM}  = \frac{1}{4} \int d^4x  F^a_{\mu\nu} F^a_{\mu\nu}   , \label{ym}
\end{equation}
$S_{GF}$ is the Landau gauge Faddeev-Popov action,
\begin{equation}
S_{GF} = \int d^{4}x  \left(
ib^{a}\,\partial_{\mu}A^{a}_{\mu}
+\bar{c}^{a}\partial_{\mu}D^{ab}_{\mu}(A)c^{b}     \right)   , \label{sfp}
\end{equation}
and finally,
\begin{eqnarray}
S_{GZ}&=&\int d^{4}x\left[\bar{\varphi}^{ac}_{\mu}\partial_{\nu}D^{ab}_{\nu}\varphi^{bc}_{\mu}-\bar{\omega}^{ac}_{\mu}\partial_{\nu}(D^{ab}_{\nu}\omega^{bc}_{\mu})-g(\partial_{\nu}\bar{\omega}^{an}_{\mu})f^{abc}D^{bm}_{\nu}c^{m}\varphi^{cn}_{\mu}\right]
\nonumber\\&&-\gamma^{2}g\int d^4 x\left[f^{abc}A^{a}_{\mu}\varphi^{bc}_{\mu}+f^{abc}A_{\mu}^{a}\bar{\varphi}_{\mu}^{bc}+\frac{d}{g}(N_{c}^{2}-1)\gamma^{2}\vphantom{\frac{1}{2}}\right],
\label{S_gamma2}
\end{eqnarray}
 is the part of the action which enables us to restrict the functional integral to $\Omega$. In expression \eqref{S_gamma2},
 $(\bar{\varphi}^{ac}_{\mu},~\varphi^{ac}_{\mu})$ is a pair of complex-conjugate bosonic fields, $(\bar{\omega}^{ac}_{\mu},~\omega^{ac}_{\mu})$ a pair of anti-commuting complex-conjugate fields, while $\gamma$ is the Gribov parameter, dynamically fixed by means of its gap equation \cite{Zwanziger:1989mf,Zwanziger:1992qr},
\begin{equation}\label{gap}
\braket{f^{abc}A_\mu^a({\varphi}_{\mu}^{bc}+\bar{\varphi}_{\mu}^{bc})}=2d(N^2-1)\frac{\gamma^2}{g^2}.
\end{equation}
 Remarkably,  the action $S$ is multiplicative renormalizable to all orders, as discussed in \cite{Zwanziger:1992qr,Maggiore:1993wq,Dudal:2010fq,Capri:2015mna}.

 After the formulation of the Gribov-Zwanziger action, it was realized that it is plagued by non-perturbative dynamical instabilities, caused by the formation of the dimension two condensates, $\langle A_{\mu}^{a}A_{\mu}^{a}\rangle$ and $\langle \bar{\varphi}^{ab}_{\mu}\varphi^{ab}_{\mu}-\bar{\omega}^{ab}_{\mu}\omega^{ab}_{\mu} \rangle$, which are energetically favoured \cite{Dudal:2007cw,Dudal:2008sp,Dudal:2011gd}.  These condensates can be taken into account from the beginning, leading to the so-called Refined-Gribov-Zwanziger action \cite{Dudal:2007cw,Dudal:2008sp,Dudal:2011gd}
\begin{equation}
S_{R}  =  S_{YM}+S_{GF}+S_{RGZ} \;, \label{sref}
\end{equation}
where
\begin{equation}
S_{RGZ}=S_{GZ}+\frac{m^2}{2}\int d^4 x A_{\mu}^{a}A_{\mu}^{a}+M^{2}\int d^4 x(\bar{\varphi}^{ab}_{\mu}\varphi^{ab}_{\mu}-\bar{\omega}^{ab}_{\mu}\omega^{ab}_{\mu}).
\label{RGZ_action}
\end{equation}
As much as the Gribov mass $\gamma^2$,  the new  parameters $(m^2, M^2)$ are dynamically determined by their own gap equations \cite{Dudal:2007cw,Dudal:2008sp,Dudal:2011gd}. As the action $S$, the refined action $S_{R}$ is multiplicative renormalizable to all orders \cite{Dudal:2007cw,Dudal:2008sp,Dudal:2011gd}.  Moreover,  the introduction of the aforementioned  condensates allows for a nice agreement with the lattice data, see e.g.~\cite{Dudal:2010tf,Cucchieri:2011ig,Dudal:2012zx}.

Recently, the Gribov-Zwanziger formalism was generalized to the linear covariant gauges \cite{Capri:2015ixa}. Simultaneously,  an exact BRST invariance of the Landau gauge actions \eqref{S_gamma}, \eqref{sref} was found, with immediate extension to the linear covariant gauges \cite{Capri:2015ixa}.  Following \cite{Capri:2015ixa},  the action \eqref{S_gamma} is replaced by
\begin{equation}
S_{\rm lc} = S_{YM} + S_{GF} + S_{RGZ} + S_{\tau}  , \label{stact}
\end{equation}
where $S_{GF}$ now denotes the Faddeev-Popov gauge-fixing in linear covariant gauges, i.e.
\begin{equation}
S_{GF} = \int d^{4}x  \left( \frac{\alpha}{2}\,b^{a}b^{a}
+ib^{a}\,\partial_{\mu}A^{a}_{\mu}
+\bar{c}^{a}\partial_{\mu}D^{ab}_{\mu}(A)c^{b}     \right)   , \label{sfpb}
\end{equation}
 with $\alpha$ denoting the gauge parameter.  The value  $\alpha=0$ corresponds to the Landau gauge.  For the Refined-Gribov-Zwanziger action in linear covariant gauges we have \cite{Capri:2015ixa}
\begin{eqnarray}\label{lact}
S_{RGZ}  &=& \int d^{4}x  \, \left(
 - \bar\varphi^{ac}_{\nu}{\cal M}^{ab}(A^h)\varphi^{bc}_{\nu}
+\bar\omega^{ac}_{\nu}{\cal M}^{ab}(A^h)\omega^{bc}_{\nu}
+\gamma^{2}g\,f^{abc}(A^h)^{a}_{\mu}(\varphi^{bc}_{\mu}+\bar\varphi^{bc}_{\mu})+\frac{m^2}{2}\int d^4 x (A^h)_{\mu}^{a}(A^h)_{\mu}^{a}\right.\nonumber\\&&+\left.M^{2}\int d^4 x(\bar{\varphi}^{ab}_{\mu}\varphi^{ab}_{\mu}-\bar{\omega}^{ab}_{\mu}\omega^{ab}_{\mu}) -d(N^2-1)\gamma^4 \right),
\end{eqnarray}
and ${\cal M}^{ab}(A^h)$ is the Hermitian, gauge invariant operator
\begin{equation}
{\cal M}^{ab}(A^h) = - \delta^{ab} \partial^2 + g f^{abc} (A^h)^c_\mu  \partial_\mu  .  \label{fpop}
\end{equation}
The configuration $A_\mu^h$ is a non-local power series in the gauge field, obtained by minimizing the functional $f_A[u]$ along the gauge
orbit of $A_{\mu }$ \cite{Dell'Antonio:1989jn,vanBaal:1991zw,Lavelle:1995ty}, with
\begin{eqnarray}
f_A[u] &\equiv &\min_{\{u\}}\mathrm{Tr}\int d^{4}x\,A_{\mu
}^{u}A_{\mu }^{u},
\nonumber \\
A_{\mu }^{u} &=&u^{\dagger }A_{\mu }u+\frac{i}{g}u^{\dagger }\partial _{\mu
}u.  \label{Aminn0}
\end{eqnarray}
One finds that a local minimum is given by
\begin{eqnarray}
A_{\mu }^{h} &=&\left( \delta _{\mu \nu }-\frac{\partial _{\mu }\partial
_{\nu }}{\partial ^{2}}\right) \phi _{\nu }\;,  \qquad  \partial_\mu A^h_\mu= 0 \;, \nonumber \\
\phi _{\nu } &=&A_{\nu }-ig\left[ \frac{1}{\partial ^{2}}\partial A,A_{\nu
}\right] +\frac{ig}{2}\left[ \frac{1}{\partial ^{2}}\partial A,\partial
_{\nu }\frac{1}{\partial ^{2}}\partial A\right] +O(A^{3}).  \label{min0}
\end{eqnarray}
It can be checked that the quantity $A_\mu^h$ is gauge invariant  order by order \cite{Capri:2015ixa}. To get control on the renormalization properties of this seemingly highly non-local quantum field theory, an equivalent local formulation can be obtained. Following \cite{Capri:2015ixa,Fiorentini:2016rwx}, we   set
\begin{equation}
 A^{h}_{\mu}=(A^{h})^{a}_{\mu}T^{a}=h^{\dagger}A^{a}_{\mu}T^{a}h+\frac{i}{g}\,h^{\dagger}\partial_{\mu}h,    \label{st}
\end{equation}
while
\begin{equation}
h=e^{ig\,\xi^{a}T^{a}},
\label{hxi}
\end{equation}
with the role of the fields $\xi^a$ akin to that  of the  Stueckelberg formulation. The local gauge invariance  of $A^{h}_{\mu}$ under a gauge transformation $u\in SU(N)$ is now immediately clear from
\begin{equation}
h\to u^\dagger h\,,\qquad\ h^\dagger\to h^\dagger u\,,\qquad A_\mu \to u^\dagger A_\mu u + \frac{i}{g}u^\dagger \p_\mu u.
\end{equation}
The   term
\begin{equation}
S_{\tau} = \int d^4x\; \tau^{a}\,\partial_{\mu}(A^h)^{a}_{\mu}   \label{stau}
\end{equation}
implements, through the Lagrange multiplier $\tau$, the transversality of the composite operator $(A^h)_\mu^a$,  namely $\partial_{\mu}(A^h)^{a}_{\mu}=0$. Solving the latter constraint  gives back the non-local
expression for the field $A^h_\mu$ expressed in \eqref{min0}. This constraint also plays a crucial role to maintain the ultraviolet renormalizability of the theory \cite{Fiorentini:2016rwx}, in sharp contrast with the standard Stueckelberg  formulation. It can also be shown that at the practical level, \eqref{S_gamma} and \eqref{stact} give rise to identical dynamics when working in the Landau gauge,  due to the gauge condition $\p_\mu A_\mu^a=0$ \cite{Capri:2015ixa}.

The action  $S_{\rm lc}$ enjoys an exact nilpotent BRST invariance,  $s S_{lc} = 0$,  expressed by \cite{Capri:2015ixa}
\begin{eqnarray}
s A^{a}_{\mu}&=&-D^{ab}_{\mu}c^{b}\,,\;\;\;\; s c^{a}=\frac{g}{2}f^{abc}c^{b}c^{c}\,,\nonumber\\
s \bar{c}^{a}&=&ib^{a}\,,\;\;\;\;
s b^{a}= 0\,,  \nonumber\\
s h^{ij} &=& -ig c^a (T^a)^{ik} h^{kj}  \;, \nonumber\\
s \varphi^{ab}_{\mu}&=& 0 \,,\;\;\;\; s \omega^{ab}_{\mu}=0\,,\nonumber\\
s\bar\omega^{ab}_{\mu}&=&0  \,,\;\;\;\; s\bar\varphi^{ab}_{\mu}=0\,,\nonumber\\
s\tau^{a}&=&0.
 \label{brstgamma}
\end{eqnarray}
A detailed analysis of the consequences of this BRST invariance at the quantum level can be found in \cite{Fiorentini:2016rwx}, where e.g.~the Nielsen identities were discussed. Notice that this BRST transformation is different from the one adopted before in e.g.~\cite{Zwanziger:1992qr}. In particular, the auxiliary fields $(\varphi^{ab}_{\mu},\bar\varphi^{ab}_{\mu},  \omega^{ab}_{\mu}, \bar\omega^{ab}_{\mu})$ are now BRST singlets, i.e.~they are BRST invariant fields. Though, as shown in details in \cite{Capri:2017bfd} they possess their own Ward identities which ensure the all orders renormalizability of the action $S_{\rm lc}$, eq.\eqref{stact}. The gap equation \eqref{gap} gets replaced by its gauge invariant counterpart $\braket{f^{abc}(A^h)_\mu^a({\varphi}_{\mu}^{bc}+\bar{\varphi}_{\mu}^{bc})}=2d(N^2-1)\frac{\gamma^2}{g^2}$. For the current work, we are mostly interested in the general form of the gluon propagator \cite{Fiorentini:2016rwx}, which can be proven to be
\begin{equation}
D_{\mu\nu}(p)=D(p)P_{\mu\nu}(p)+L(p)\frac{p_{\mu}p_{\nu}}{p^{2}},
\label{gluon_prop_faynman}
\end{equation}
with the transverse form factor,
\begin{eqnarray}\label{rgz}
  D(p)&=&\frac{p^2+M^2}{p^4+(M^2+m^2)p^2+M^2m^2+\lambda^4},
\end{eqnarray}
containing all non-trivial information.
 At tree level, this factor stems from the quadratic part of the action \eqref{stact}, where we set $\lambda^4=2g^2N\gamma^4$.
Next to
\begin{equation}
L(p)=\frac{\alpha}{p^2},
\label{gluon_prop_faynmanbis}
\end{equation}
with
\begin{equation}
P_{\mu\nu}(p)=\delta_{\mu\nu}-\frac{p_{\mu}p_{\nu}}{p^{2}}\,,\qquad L_{\mu\nu}(p)=\frac{p_{\mu}p_{\nu}}{p^{2}}
\label{tranverse_prop}
\end{equation}
the transversal and longitudinal projectors. Analogously as in perturbation theory, the longitudinal gluon propagator is exactly known, being equal to its tree level expression, as it follows from the BRST symmetry \eqref{brstgamma}.   The same behavior for the longitudinal component of the propagator is found in other non-perturbative  approaches to the linear covariant gauges  \cite{Aguilar:2015nqa}.


 It is worth mentioning here that expression \eqref{rgz} fits very well the lattice data, see ~\cite{Dudal:2010tf,Cucchieri:2011ig,Dudal:2012zx}. Moreover, when the various mass parameters are estimated by means of the direct lattice comparison ~\cite{Dudal:2010tf,Cucchieri:2011ig,Dudal:2012zx}, it turns out that expression \eqref{rgz}  exhibits  2 complex-conjugate poles, indicating that the corresponding degrees of freedom cannot be  associated to  excitations of the physical spectrum, similar to the rationale of e.g.~\cite{Kharzeev:2015xsa} or \cite{Hawes:1993ef}. Said otherwise, expression \eqref{rgz}  can be seen as a manifestation of gluon confinement. Propagators with complex poles such as \eqref{rgz} have also been used to fit finite temperature lattice data \cite{Aouane:2011fv}, based on which the confinement-deconfinement transition was discussed using the Polyakov loop criterion, providing a reasonable estimate for the critical temperature, see e.g.~\cite{Fukushima:2012qa,Canfora:2015yia}. How to connect the (infrared) behaviour of the gluon, quark and ghost propagators and their mutual interaction vertices to the linear confining potential between static charges etc.~is an open question, for whatever analytical scheme is considered to study the elementary $n$-point functions of QCD.

\section{ The topological susceptibility}
Returning to QCD with 3 light flavours $(u, d, s)$, we recall it  enjoys an (almost) $U_L(3)\times U_R(3)$ (left $\times$ right) symmetry. There is a dynamical chiral (axial~$=  L-R$) symmetry breaking, reducing the invariance to the vector part ($L+R$), $U_V(3)$. So one experimentally expects a nonet of (almost) massless Goldstone modes, viz~ the 3 pions, 4 kaons, $\eta$ and $\eta'$. Though, it turns out that the $\eta'$ particle is way too massive to be called the ninth ``almost Goldstone'' boson as $m_{\eta'}\approx 958~\text{MeV}$.

One solution to this apparent $U_A(1)$ problem was offered by 't Hooft by means of instanton calculus, see \cite{tHooft:1976rip}. The validity of the instanton calculus in the large $N$ limit is a delicate issue (see \cite{marino} for a recent account), but another way to understand the anomalous $\eta'$ mass was, independently, worked out by Veneziano and Witten in \cite{Witten:1979vv,Veneziano:1979ec} \footnote{Due to the intricate infrared problems accompanying instantons in an infinite volume, a large $N$ extrapolation is difficult when not working in a finite volume \cite{marino}}. Summarizing, the celebrated Veneziano-Witten formula reads
\begin{equation}\label{wz}
m_{\eta'}^2=\frac{4N_f}{f_\pi^2}\chi^4_{\theta=0,N_f=0}=\mathcal{O}(1/N),
\end{equation}
where $\theta$ is the vacuum angle and $f_\pi$ the pion decay constant. Although simple at first sight, this is a very intricate formula, since the l.h.s.~refers to QCD (with $N_f$ flavours), while the
r.h.s.~to the pure gauge theory. The relation \eqref{wz} thus explains the relatively large $\eta'$ mass, given that $\chi^4_{\theta=0,N_f=0}\equiv \chi^4$ is sufficiently large. Filling in the numbers requires $\chi^4\sim (200~\text{MeV})^4$, not far from the lattice $SU(3)$ estimates as reported in \cite{Alles:1996nm}.

Returning to the work \cite{Kharzeev:2015xsa}, it was suggested that a resolution of the Gribov copies problem, being intimately linked with the topology of the gauge group, might also be of direct importance to capture the non-trivial topological structure of the QCD vacuum. In Euclidean space-time, we have the classical instanton solutions, describing in Minkowski space-time the tunneling between the degenerate vacuum states with different Chern-Simons charge \cite{Kharzeev:2015ifa},
\begin{equation}
X=\int d^3 xK_0,
\label{CS_number}
\end{equation}
with $K_0$ the temporal component of topological Chern-Simons current,
\begin{equation}
K_\mu =\frac{g^2}{16\pi^2}\epsilon_{\mu\nu\rho\sigma}A_{\nu,a}\left(\partial^\rho A^{\sigma,a}+\frac{g}{3}f^{abc}A^\rho _b A^\sigma _c\right).
\label{CS_current}
\end{equation}
This current is related to the topological charge density,
 \begin{equation}
 Q(x)=\partial_\mu K_\mu=\frac{g^2}{32\pi^2}F_{\mu\nu}\Tilde{F}_{\mu\nu}.
 \label{top_charge_density}
 \end{equation}
Witten and Veneziano suggested that the vacuum topology fluctuations can be captured by the occurrence of an unphysical mass pole \cite{Witten:1979vv,Veneziano:1979ec}, the Veneziano ghost, in the topological current correlator,
\begin{equation}
p_{\mu}p_{\nu}\Braket{K_{\mu}K_{\nu}}_{p=0}\neq 0.
\label{KK}
\end{equation}
Thus, the Veneziano solution was to assume that
\begin{equation}
K_{\mu\nu}(p)=i\int d^{4}x\  e^{ipx}\left<K_{\mu}(x)K_{\nu}(0)\right>\stackrel{p^{2}\thicksim 0}\thicksim -\frac{\chi^{4}}{p^{2}}g_{\mu\nu},
\label{vene_prop.}
\end{equation}
where $\chi^4\geq0$ is the topological susceptibility of pure Yang-Mills theory. The negative sign in \eqref{vene_prop.} means that we are dealing with an unphysical ghost particle, so it cannot be directly measured in a physical process, though the couplings of the ghosts can influence physical amplitudes \cite{Kharzeev:2015xsa}.

Kharzeev and Levin \cite{Kharzeev:2015xsa,Kharzeev:2015ifa} interpreted the current correlator \eqref{vene_prop.} as resulting from an effective interaction between the gluon (in Feynman gauge $\alpha=1$) and the Veneziano ghost \cite{Kharzeev:2015xsa,Kharzeev:2015ifa,Dudal:2015khv}. An effective ghost-gluon vertex $\Gamma_{\mu}(q,p)$ was postulated, and then it was found that a dynamically corrected gluon propagator (the ``glost''),
\begin{equation}
\mathcal{G}_{\mu\nu}(p^{2})=\frac{p^{2}}{p^{4}+\chi^{4}}\delta_{\mu\nu},
\label{glost}
\end{equation}
solves the Dyson-Schwinger equation, when using only this coupling \cite{Kharzeev:2015xsa,Dudal:2015khv} in the deep infrared. Immediately, we notice that there is an inconsistency between \eqref{gluon_prop_faynman} and \eqref{glost}, indicating that the propagator \eqref{glost} is incompatible with BRST symmetry  as well as with all independently obtained functional forms for the non-perturbative Feynman gauge gluon propagator, see also \cite{Dudal:2015khv} for more concerns concerning the construction proposed in \cite{Kharzeev:2015xsa}.

\section{Setup of a rational (Pad\'{e}) approximation via the spectral K\"all\'{e}n-Lehmann representation of the topological current correlator}
From now on, we wish to find out if, by using Gribov type propagators, we can obtain a reasonable ``semi-non-perturbative'' estimate for the topological susceptibility $\chi^4$, without the need to introduce new effective vertices.  We notice that transversal  component $D(p)$ of the propagator in \eqref{rgz} can be written as the linear combination of 2 standard massive propagators with complex-conjugate masses, which allows for standard Feynman diagram-computational manipulations.

We will follow the Euclidean conventions of \cite{Meggiolaro:1998bh}. More precisely, we have
\begin{equation}\label{v2b}
  \chi^4 = -\lim_{p^2\to0} p_\mu p_\nu \braket{K_\mu K_\nu}\geq0.
\end{equation}
That $\chi^4\geq 0$  follows from translational invariance applied to
\begin{equation}\chi^4 = \int d^4x \braket{F\tilde F(x),F\tilde F(0)}=V^{-1}\int d^4x \int d^4y \braket{F\tilde F(x),F\tilde F(y)}\geq 0.\end{equation}

As pointed out originally in \cite{Seiler:1987ig,Seiler:2001je}, see also \cite{Vicari:2008jw}, it holds that
\begin{equation}\label{rf1}\braket{Q(x)Q(0)}<0\,,\qquad\text{for~} |x|>0,\end{equation} due to the (Osterw\"alder-Schrader) reflection positivity  and to  the $t$-odd character of $Q$. Combined with $\int d^4x \braket{Q(x)Q(0)}>0$, this entails that $\braket{Q(x)Q(0)}$ must contain a positive contact term to compensate the negative rest of the integral. For example, setting
\begin{equation}\mathcal{Q}(x)=\braket{Q(x)Q(0)}\,,\qquad\text{for~} |x|>0,\end{equation} then we need
\begin{equation}\braket{Q(x)Q(0)}=\mathcal{Q}(x) +C\delta(x)\,,\qquad\text{for~} |x|\geq0,\end{equation}
with $C>0$.

The contact term thus plays a pretty important role in the \emph{definition} of the topological susceptibility, as recognized in \cite{Seiler:1987ig}.
Contact terms are evidently also important to get a finite (cf.~additive renormalization) value for $\braket{Q(x)Q(0)}$. As in principle these contact terms can be chosen, one might have the impression that this would reflect in a randomness in the definition of $\chi^4$. This is however not the case, see later.

Closely related to the above comment is that the $t$-odd character of $Q$ also means that the K\"all\'{e}n-Lehmann spectral density of $\braket{QQ}$ is negative. The original remarks can be found in \cite{Seiler:1987ig,Seiler:2001je}. Following the derivation of the usual Minkowski 2-point function of $Q$, we will find a positive spectral integral. Though, upon passing to Euclidean space-time via a Wick rotation, the $t$-odd nature of $Q$ will  introduce an extra factor  $i$ \footnote{Compare indeed to the complex $i\theta F\tilde F$ topological term in the Euclidean YM action. }, meaning that $\braket{QQ}$ in Euclidean spacetime will pick up an extra overall minus. Thus, actually, we have
        \begin{equation}\label{KLQ}\hat{\mathcal{Q}}(p^2)\equiv\braket{Q(p)Q(-p)}=-\int_0^\infty d\mu\frac{\rho(\mu)}{\mu+p^2}\,,\qquad \rho(\mu)\geq 0.\end{equation}

It is interesting to remark that \eqref{KLQ} is consistent with \eqref{rf1} when we think in terms of the usual temporal Schwinger function that is used to check for positivity violations \cite{Alkofer:2003jj}. Indeed, usually we introduce, for $t>0$,
        \begin{equation}\label{temps}C(t)=\frac{1}{2\pi} \int_{-\infty}^{\infty} dp e^{-ipt} \hat{\mathcal{Q}}(p^2).\end{equation}
    Substituting \eqref{KLQ} into \eqref{temps}, we can rewrite this as
        \begin{equation}\label{temps2}C(t)=-\frac{1}{2} \int_{0}^{\infty}d\mu\frac{\rho(\mu)}{\sqrt{\mu}}e^{-t\sqrt{\mu}}= - \int_{0}^{\infty}d\nu\rho(\nu^2)e^{-t\nu}<0.\end{equation}
    The definition \eqref{temps} can however be rewritten as, with $|x|=|(t,\vec{x})|>0$,
        \begin{equation}\label{temps3}C(t)=\frac{1}{2\pi}\frac{1}{V_3} \int d^3x\int d^4p e^{-ipx} \hat{\mathcal{Q}}(p^2).\end{equation}
        We recognize the Fourier transform of $\mathcal{Q}(x)$, so we recover
        \begin{equation}\mathcal{Q}(t)\propto C(t)<0,
        \end{equation}
        i.e.~, the negative spectral density in the K\"all\'{e}n-Lehmann representation \eqref{KLQ} indeed ensures \eqref{rf1}.

Let us now carry out a power counting analysis. As $\dim\hat{\mathcal{Q}}(p^2)=4$, this correlation function needs 3 subtractions to make it well defined (finite), as $\rho(\tau)\sim \tau^2$ for $\tau\to\infty$. Without loss of generality, we can subtract at zero momentum, finding as proper version of \eqref{KLQ}
\begin{equation}\label{KLQsub}\hat{\mathcal{Q}}(p^2)=a_0+a_1 p^2+ a_2p^4-p^6\int_0^\infty d\tau\frac{\rho(\tau)}{(\tau+p^2)\tau^3}\end{equation}

Doing so, $\chi^4$ would be determined by the subtraction constant $a_0$. This is a somewhat undesirable feature, as we wish to derive an estimate for $\chi^4$ from the correlator. Though, usually one employs  the a priori knowledge of the correlator,   and possibly  of its slope, at zero momentum via low energy theorems, to exchange constants like $a_0$, $a_1$ in terms of  these a priori known numbers (like $\chi^4$) to completely fix the correlation at any momentum. This is rather the  inverse order of the current analysis, where $\chi^4$ is  unknown. For a few examples we refer to \cite{Narison:2002pw}.

Next, including quarks in the analysis, it was shown in \cite{Crewther:1978kq} that the susceptibility, as defined via \eqref{v2b}, is the one entering the anomalous chiral Ward identities. As discussed in \cite{Meggiolaro:1998bh}, this definition also concurs with the one given in \cite{Witten:1979vv} via the double derivative of the $\theta$ term in the action.

Let us show, employing expression \eqref{v2b},   also removes any ambiguity imposed by the subtraction procedure. We may in general set
\begin{eqnarray}\label{k1}
  \braket{K_\mu(p) K_\nu(-p)}&=&\left(\delta_{\mu\nu}-\frac{p_\mu p_\nu}{p^2}\right)\mathcal{K}_\perp(p^2)+\frac{p_\mu p_\nu}{p^2}\mathcal{K}_\|(p^2)\nonumber\\&\equiv&\left(\delta_{\mu\nu}-\frac{p_\mu p_\nu}{p^2}\right)\int_0^\infty  d\tau\frac{\rho_\perp(\tau)}{\tau+p^2}+\frac{p_\mu p_\nu}{p^2}\int_0^\infty d\tau \frac{\rho_\|(\tau)}{\tau+p^2},
\end{eqnarray}
based on Euclidean invariance. Then, we already find that\footnote{We temporarily ignored the necessary subtractions here, see later.}
\begin{equation}\label{k2}
  \hat{\mathcal{Q}}(p^2)=-p^2\mathcal{K}_\|(p^2)=-p^2\int_0^\infty d\tau \frac{\rho_\|(\tau)}{\tau+p^2}
\end{equation}
and thus
\begin{equation}\label{k3}
  -\chi^4=\lim_{p^2\to0}p^2\mathcal{K}_\|(p^2)=\lim_{p^2\to0}p^2\int_0^\infty d\tau \frac{\rho_\|(\tau)}{\tau+p^2}.
\end{equation}
As the l.h.s.~of \eqref{k2} is gauge invariant, so should the r.h.s.~be, meaning that the longitudinal form factor $\mathcal{K}_\|(p^2)$ and its associated  spectral function $\rho_\|(\mu)$ ought to be gauge invariant. Likewise, the transversal piece may contain gauge variant  contributions  depending on the gauge parameter $\alpha$. This is the reason why we opt to work with a general linear covariant gauge, as we can then explicitly check how the gauge (in)variance manifests itself in the full $\braket{KK}$-correlation function.

Let us now be a bit more careful, and include the subtraction terms. We focus on the  relevant longitudinal sector. From dimensional analysis, it is clear that this time we only need 2 subtractions ($\rho_\|(\tau)\sim \tau$ for $\tau\to\infty$), so a finite result\footnote{It has been shown that the topological charge operator itself is a renormalization group invariant \cite{Vicari:2008jw,Espriu:1982bw} in pure gauge theories (it is not when dynamical quarks are included,  in which case even operator mixing occurs with the chiral current). Though we are interested in the topological susceptibility of the pure gauge theory.  Notice that it is indeed this quantity which enters the Witten-Veneziano formula for the $\eta'$ mass \cite{Witten:1979vv,Veneziano:1979ec}.  As a consequence, the leading ultraviolet behaviour in the $p^2\to\infty$ limit of the correlator $\mathcal{K}_\|(p^2)$ will also be completely determined by its tree level behaviour, taking into account the asymptotic freedom and vanishing anomalous dimension of the topological charge operator.} is guaranteed from
\begin{equation}\label{k7}
  \mathcal{K}_\|(p^2)=b_0+b_1p^2+p^4\int_0^\infty d\tau \frac{\rho_\|(\tau)}{(\tau+p^2)\tau^2}
\end{equation}
and thus
\begin{equation}\label{k4}
  -\chi^4=\lim_{p^2\to0}p^2\left(b_0+b_1p^2+p^4\int_0^\infty d\tau \frac{\rho_\|(\tau)}{(\tau+p^2)\tau^2}\right),
\end{equation}
with $b_{0,1}$ subtraction constants. Obviously, we can rewrite \eqref{k4} as
\begin{equation}\label{k5}
  -\chi^4=\lim_{p^2\to0}p^6\int_0^\infty d\tau \frac{\rho_\|(\tau)}{(\tau+p^2)\tau^2}
\end{equation}
and all reference to subtraction constants is indeed gone.

In a loop expansion approach using the RGZ action and associated RGZ propagator \eqref{rgz}, \eqref{k5} will still automatically vanish. It is important to stress here how difficult it is to get non-zero value for the topological susceptibility from a continuum viewpoint. Obviously, in the absence of a dynamical singularity at $p^2\to0$ in the correlation function of two topological currents, $\chi^4$ will always automatically vanish. This is precisely why one needs a dynamically generated massless (ghost) bound state in the $K_\mu$ channel.   Usually, to get an estimate for a bound state mass, one makes a series of approximations and/or assumptions and obtain an estimate for e.g. the mass of a pion.  Here, an approximation of the mass of the Veneziano ghost would be insufficient, as we need it to be \emph{exactly} zero.
Usually, a massless bound state can be probed when a symmetry (and symmetry-consistent approximations) protects the mass to be zero, needless to say we are thinking about the Goldstone mechanism here. Clearly, the latter does not apply to the Veneziano ghost particle, which is not the Goldstone degree of freedom created by a current corresponding to a spontaneously broken global symmetry. On the contrary, it is a topologically conserved current.
The inherent difficulty just sketched to get a continuum handle on the Veneziano ghost is what, to our understanding, explains the lack of references trying to tackle the problem directly.  Notice that this does not a priori exclude a truly non-perturbative lattice study of the relevant correlator, apart from the potential difficulties in precisely defining the topological current on the lattice \cite{Luscher:1981zq,Seiberg:1984id,Karsch:1993fv,Luscher:2004fu}.

Therefore, in this section we work out an approximation to the quantity appearing in \eqref{k5}, more precisely to
\begin{equation}\label{nieuwevgl1}
  p^6\int_0^\infty d\tau \frac{\rho_\|(\tau)}{(\tau+p^2)\tau^2}.
\end{equation}
Evidently, e.g.~the one loop approximation to $\mathcal{K}_\|(p^2)$ is not exact, since RGZ is still meant to be an expansion in the YM coupling $g^2$, though on top of a nontrivial vacuum, encoded in the condensates,  e.g.~the dynamical mass scales,  present in \eqref{rgz}. Due to asymptotic freedom, the one loop result will be trustworthy at sufficiently large $p^2$. As such, we will only retain the leading $\frac{1}{p^2}$ power correction (proportional to the mass scales present in the RGZ propagator) relative to  the standard perturbative one loop estimate for $\mathcal{K}_\|(p^2)$.

Assuming we temporarily rewrite the RGZ gluon propagator as
\begin{equation}
D(p^{2})=\frac{p^{2}+M_{1}^2}{p^4+M_2^2p^2+M_3^4},
\label{Prop_gluon}
\end{equation}
we first need to extract the spectral density associated with the K\"all\'en-Lehmann representation of the physical part of the $K_\mu$ correlation function. This has been worked out in full detail in the Appendix \ref{appA}, leading to \eqref{rho_tau},
\begin{equation}\label{herhaling}
  \rho_\|(\tau)=-2A_+A_-\frac{g^{4}(N^{2}-1)}{2^{2d+5}\pi^{7/2}\Gamma(\frac{d-1}{2})}\frac{\left(\tau^{2}-4b^{2}-4a\tau\right)^{(d-1)/2}}{\tau^{d/2}}\qquad\text{for}~\tau\geq \tau_c=2(a+\sqrt{a^2+b^2}),
\end{equation}
where
\begin{equation}\label{herhaling2}
  a=\frac{M_2^2}{2}\,,\qquad b=\frac{\sqrt{4M_3^4-M_2^4}}{2}.
\end{equation}
 A formal way to derive the spectral density based on properties of the Stieltjes transform can be found in \cite{Dudal:2010wn}. The threshold $\tau_c$ is the natural generalization to the case of 2 complex conjugate masses of the usual threshold appearing at $(m+m')^2$ in the case of 2 standard particles with masses $m$ and $m'$.

It is perhaps interesting to notice here that other non-perturbative approaches do also give reasonably good descriptions of the lattice data, see f.i.~\cite{Aguilar:2008xm,Fischer:2008uz,Cyrol:2016tym,Serreau:2012cg}, but in particular is the output of functional equations as used in \cite{Aguilar:2008xm,Fischer:2008uz,Cyrol:2016tym} purely numerical, which would make the determination of the here required spectral density much more cumbersome, perhaps possibly via the (rather involved) numerical routine of \cite{Windisch:2012zd}. Therefore, we will solely rely here on the discussed RGZ propagator, which allows for closed analytical expressions.

As we use lattice data to get an estimate of the parameters $M_i$, we need to work in a lattice compatible renormalization scheme, such as the  momentum subtraction MOM scheme, defined by
\begin{equation}
D(p^{2}=\mu^2)=\frac{1}{\mu^2}.
\label{prop_gluon_MOM}
\end{equation}
The proper renormalization factor $Z$,   at scale $\mu$, is thus given by
\begin{equation}
D(p^{2})=Z\frac{p^{2}+M_{1}^2}{p^4+M_2^2p^2+M_3^4},
\label{Prop_gluon_MOM_scale}
\end{equation}
with
\begin{equation}
Z=\frac{1}{\mu^2}\frac{\mu^4+M^2_2 \mu^2+M_3^4}{\mu^2+M_1^2}.
\label{factor_Z}
\end{equation}
 Since the gluon propagator we will use is the one renormalized in MOM scheme at scale $\mu$,  the  coupling constant $g^2$ present in \eqref{rho_tau} becomes
\begin{equation}
g^2(\mu)=\frac{1}{\beta_{0}\log\left(\frac{\mu^2}{\Lambda_{\mbox{\tiny{MOM}}}^2}\right)} \,,\qquad \beta_0=\frac{11}{3}\frac{N}{16\pi^2}.
\label{g_mu}
\end{equation}
We have checked the conversion formulae of \cite{Boucaud:2008gn}, leading to the following relation, valid for generic $N$ and $N_f=0$ between the MOM and $\MSbar$ scales.
\begin{equation}
    \Lambda_{\mbox{\tiny{MOM}}}=\lms e^{169/264}.
\end{equation}
This relation follows from the general theory of \cite{Celmaster:1979km}, using the following relation between the $\MSbar$ ($\overline g^2$) and MOM coupling ($g^2$),
\begin{equation}
g^2=\overline g^2(1+c_1\overline g^2 + \ldots)\,,\qquad c_1=\frac{169}{36}N.
\end{equation}
For $\lms^{N=2}$, we can use the estimate of \cite{Lucini:2008vi},
\begin{equation}
\lms^{N=2}\approx 0.752~\sqrt{\sigma}
\end{equation}
with $\sigma$ the string tension to set the physical scale on a lattice. Using the standard value $\sqrt{\sigma}\approx 0.44~\text{GeV}$, we find
\begin{equation}
\lms^{N=2}\approx 331~\text{MeV}
\end{equation}
and thus
\begin{equation}
\Lambda_{\mbox{\tiny{MOM}}}^{N=2}\approx 628~\text{MeV}.
\label{Lambda_MOM_2}
\end{equation}
Moreover, \cite{Lucini:2008vi} also reports $\lms^{N=3}\approx 0.538\sqrt{\sigma}\approx 237~\text{MeV}$, a value which compares favourably well with the estimate of \cite{Boucaud:2008gn}, stating $\lms^{N=3}\approx  224~\text{MeV}$. The (somewhat older) work \cite{Boucaud:2001st} predicted $\lms^{N=3}\approx  233~\text{MeV}$. The more recent work \cite{Sternbeck:2010xu} gave a preliminary value of $\lms^{N=3}\approx  0.62r_0\approx 262~\text{MeV}$ by using $r_0\approx2.367/\text{GeV}$. Using the estimate of \cite{Boucaud:2008gn}, we get
\begin{equation}\label{MOM3}
\Lambda_{\mbox{\tiny{MOM}}}^{N=3}\approx 425~\text{MeV}.
\end{equation}
Let us first work out the $SU(3)$ case. Including the renormalization factors $Z$, the eventual spectral density can be obtained from \eqref{rho_tau},
\begin{eqnarray}
\rho_\|(\tau)=-2A_+A_-\frac{g^{4}(\mu)Z^2}{2^{9}\pi^{4}}\frac{\left(\tau^{2}-4b^{2}-4a\tau\right)^{3/2}}{\tau^{2}}.
\label{rho_tau_Z}
\end{eqnarray}
Using the lattice obtained values \cite{Oliveira:2012eh}
\begin{equation}\label{fitje}
M_{1}^2=4.473\pm0.021~\text{GeV}^2\,,\qquad M_{2}^{2}=0.704\pm0.029~\text{GeV}^2\,,\qquad M_{3}^4=0.3959\pm0.0054~\text{GeV}^4\,,
\end{equation}
we get as central values
\begin{eqnarray}
a=0.352~\text{GeV}^2\,,\qquad b=0.522~\text{GeV}^2\,,\qquad 2A_+A_-=31.719.\end{eqnarray}
The fitted estimates of the mass scales entering \eqref{rgz} were obtained by matching the tree level propagator on top of the non-perturbative gluon lattice data, as a way to determine the size of the dynamical RGZ mass scales. The latter capture the non-perturbative nature of the nontrivial RGZ vacuum, around which we expect perturbation theory to work, as the RGZ mass scales offer a dynamical screening of the Landau pole. At the to be  considered scales $\mu$, relative to the MOM scale \eqref{MOM3}, the corresponding MOM strong coupling expansion parameter is effectively very small, an indication  that a perturbative treatment certainly makes sense in the considered momentum region, after which we, using the described Pad\'{e} analysis, ``extrapolate'' to the deep infrared, where we can no longer trust our perturbative result, albeit corrected with power corrections, for the correlation function. This is also the region where we eventually have to consider the zero momentum limit to make contact with the topological susceptibility. The Pad\'{e} extrapolation is thus used to estimate the hard to access small momentum behaviour of a correlation function from its controllable behaviour at higher momentum. This shares a certain resemblance with the (Laplace or other) sum rules approaches to estimate the topological susceptibility, where the ultimate source of non-perturbative effects is also tracing back to non-trivial QCD vacuum condensates that enter the Operator Product Expansion of a specific correlation function, from which is, after transforming, then also extracted the information of interest by scanning for an optimal parameter space \cite{Narison:2002pw,Narison:1994hv}.

More precisely, we approximated \eqref{nieuwevgl1} with an $[\mathcal{M}+2,\mathcal{M}]$ Pad\'e rational function in variable $p^2$, which are the ones having the same large $p^2$ behavior, viz.~$\mathcal{O}(p^4)$. We opted to do the Pad\'e approximation around $p^2=\mu^2$. In general, given a function $f(x)$, its $[\mathcal{N},\mathcal{M}]$ Pad\'e approximant $R_{\mathcal{N},\mathcal{M}}^{x_0}(x)$ is given by
\begin{eqnarray}
  R_{\mathcal{N},\mathcal{M}}^{x_0}(x) &=& \frac{a_0+\ldots + a_{\mathcal{N}}x^{\mathcal{N}}}{1+\ldots+ b_{\mathcal{M}}x^{\mathcal{M}}}
\end{eqnarray}
such that the Taylor series around $x_0$ of $R_{\mathcal{N},\mathcal{M}}^{x_0}(x)$ and $f(x)$ coincide up to order $\mathcal{N}+\mathcal{M}$.

With this, we can study the  function $\chi(\mu^2)$ using the previous ingredients and search for optimal values, in the sense of minimal dependence, on the scale $\mu^2$. We remark here that the scale at which we do the Pad\'e approximation should be not too small, so that we can trust the (perturbatively) computed r.h.s.~of \eqref{k5}, and not too large so that we are taking into account sizable non-perturbative effects from the presence of the RGZ mass scales in \eqref{rgz}, and  perform  a sensible extrapolation of the approximant to zero momentum to get an estimate for $\chi^4$. A natural choice is to expand around the renormalization scale $\mu^2$, since the MOM renormalization scale is
subject to the same assumptions when used to renormalize lattice data, see e.g.~\cite{Boucaud:2000ey}. The results are shown in FIG.~1 for $\mathcal{M}=1,2,3$ for a reasonable interval for $\mu^2$, to be compared with the lattice ballpark value of $\chi\sim 200~$MeV \cite{Alles:1996nm}. It is not a surprise that the results are pushed down as $\cal M$ grows, since for ${\cal M}\to\infty$ the approximant will converge to the original propagator which we know to have a trivial $\chi$. So, although Pad\'{e} approximation suggests a nonzero value for $\chi$, it is difficult to provide a definite estimate.

To get an error estimation from the uncertainty on $\vec{x}\equiv(M_1^2,M_2^2, M_3^4)$, we compute the corresponding standard deviation on $\chi(\mu^2)$ in the standard way,
\begin{eqnarray}
  \sigma_\chi(\mu^2)=\sqrt{\sum_i \left(\frac{\p \chi}{\p x_i}\right)^2 \sigma_{x_i}^2},
\end{eqnarray}
as the errors on the $x_i$ are small; the $\sigma_i$ can be read off from \eqref{fitje}. We have displayed $\sigma_\chi(\mu^2)$ in Fig.~2.

\begin{figure}[h]
\includegraphics[width=0.7\textwidth]{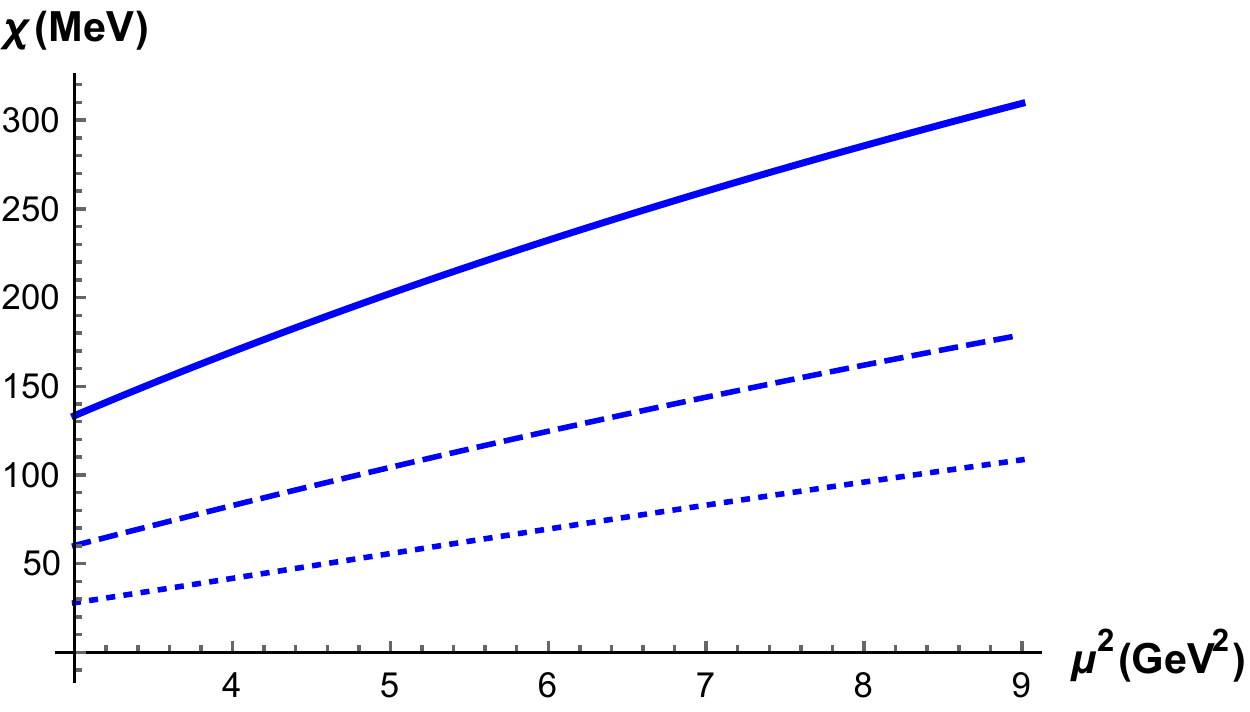}
\caption{The $SU(3)$ topological susceptibility $\chi$ for variable $\mu^2$ for $\mathcal{M}=1,2,3$ (full, dashed, dotted).}
\label{chi}
\end{figure}
\begin{figure}[h]
\includegraphics[width=0.7\textwidth]{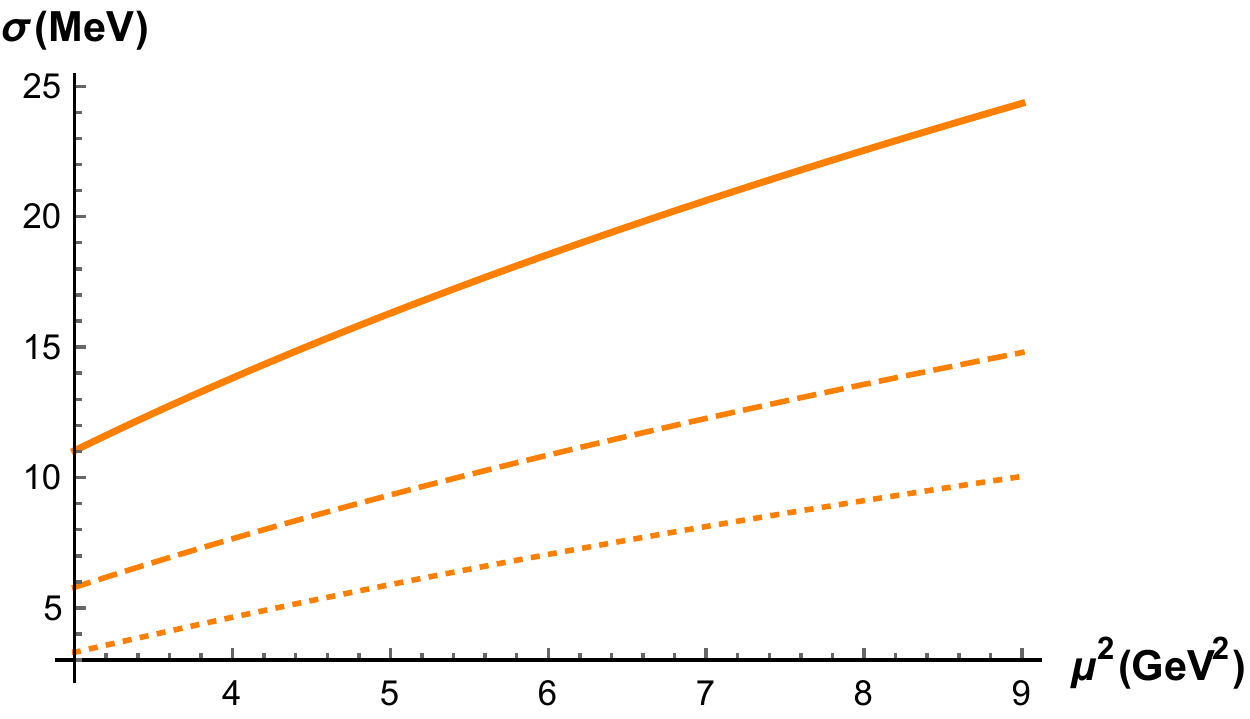}
\caption{Estimated error on $\chi$ for $SU(3)$ due to the uncertainty on the fitting parameters for variable $\mu^2$ for ${\cal M}=1,2,3$ (full, dashed, dotted).}
\label{chi}
\end{figure}

For $N=2$, $\Lambda_{\mbox{\tiny{MOM}}}$ is given by \eqref{Lambda_MOM_2} and the spectral density thence reads
\begin{eqnarray}
\rho_\|(\tau)=-2A_+A_-\frac{3g^{4}(\mu)Z^2(\mu)}{2^{12}\pi^{4}}\frac{\left(\tau^{2}-4b^{2}-4a\tau\right)^{3/2}}{\tau^{2}}.
\label{rho_tau_Z_2}
\end{eqnarray}
Following the same procedure as for $N=3$, we get the graphs of FIG.~3 and FIG.~4 in the $N=2$ case. Here,  we used
\begin{equation}
M_{1}^2=2.508\pm0.078~\text{GeV}^2\,,\qquad M_{2}^{2}=0.590\pm0.026~\text{GeV}^2\,,\qquad M_{3}^4=0.518\pm0.013~\text{GeV}^4,
\end{equation}
 as can be inferred from the largest volume data of Table II in \cite{Cucchieri:2011ig}, yielding as central values
\begin{eqnarray}
a=0.295~\text{GeV}^2\,,\qquad b=0.657~\text{GeV}^2\,,\qquad 2A_+A_-=6.176.\end{eqnarray}

\begin{figure}[h]
\includegraphics[width=0.7\textwidth]{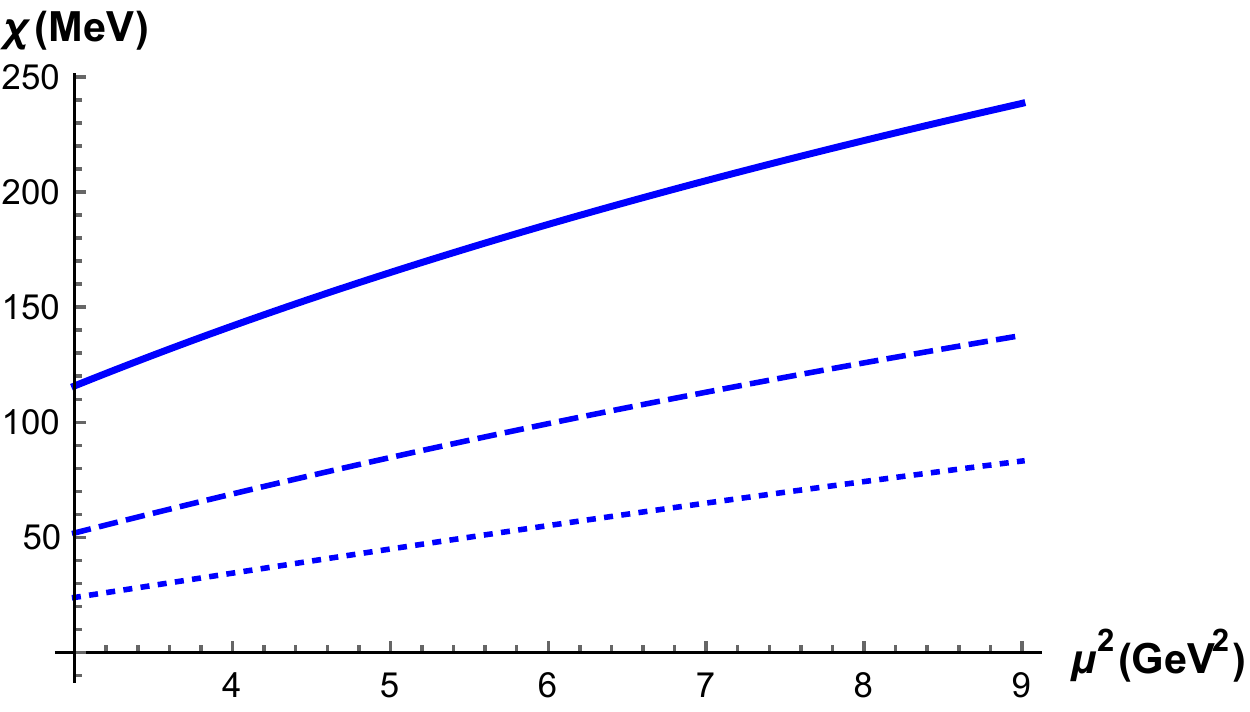}
\caption{The $SU(2)$ topological susceptibility $\chi$ for variable $\mu^2$ for ${\cal M}=1,2,3$ (full, dashed, dotted).}
\label{chi}
\end{figure}
\begin{figure}[h]
\includegraphics[width=0.7\textwidth]{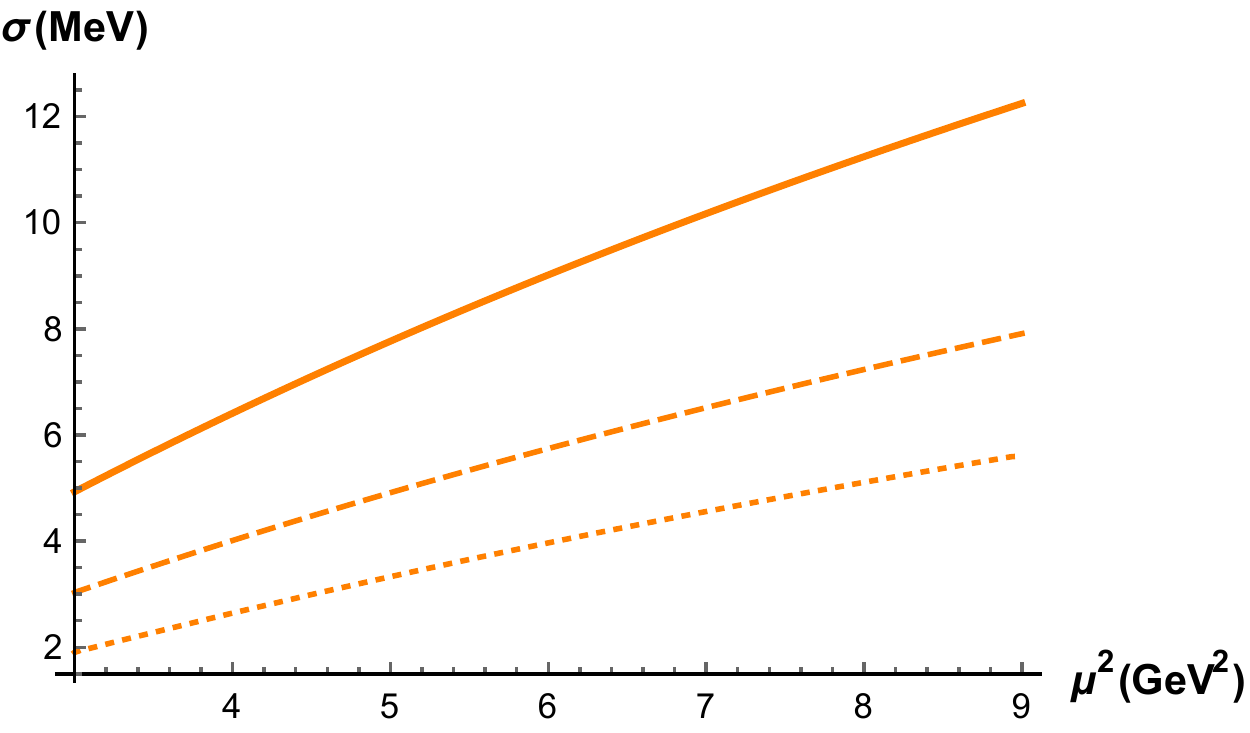}
\caption{Estimated error on $\chi$ for $SU(2)$ due to the uncertainty on the fitting parameters for variable $\mu^2$ for ${\cal M}=1,2,3$ (full, dashed, dotted).}
\label{chi}
\end{figure}

For the record, let us mention that the $SU(2)$ lattice prediction for the topological susceptibility  sets $\chi=200-230~\text{MeV}$, see \cite{Teper:1999wp}. A compatible value was also found in \cite{Campagnari:2008yg} using a non-perturbative continuum Hamiltonian approach in Coulomb gauge.

\section{Conclusion}
We have analyzed the topological susceptibility, $\chi^4$, in $SU(2)$ and $SU(3)$ Euclidean Yang-Mills theory in a generic linear covariant gauge taking into account the Gribov ambiguity. We employed a recently  constructed effective action that implements a restriction of the gauge field path integration to a suitable subregion so that at least the infinitesimal gauge copies are eliminated, this without violating the BRST symmetry. As a consequence, the topological susceptibility is, as required, gauge invariant in this non-perturbative framework, explicitly checked to leading order in the  present  work. In an attempt to get estimates for the topological susceptibility, we developed a particular Pad\'{e} rational function approximation based on the K\"all\'en-Lehmann spectral integral representation of the topological current correlation function. To improve upon the presented crude estimates, we plan to include the next order correction in future work. Notice this will be computationally challenging, thanks to the significantly enlarged set of vertices in the now considered  Refined Gribov-Zwanziger action for the linear covariant gauge.

\section*{Acknowledgments}
The Conselho Nacional de Desenvolvimento Cient\'{i}fico e
Tecnol\'{o}gico (CNPq-Brazil), the Faperj, Funda{\c{c}}{\~{a}}o de
Amparo {\`{a}} Pesquisa do Estado do Rio de Janeiro, the SR2-UERJ
and the Coordena{\c{c}}{\~{a}}o de Aperfei{\c{c}}oamento de
Pessoal de N{\'\i}vel Superior (CAPES) are gratefully acknowledged
for financial support. S.~P.~Sorella is a level PQ-1 researcher under the program Produtividade em Pesquisa-CNPq, 300698/2009-7; M.~S.~Guimaraes is supported by the Jovem Cientista do Nosso Estado program - FAPERJ E-26/202.844/2015, is a level PQ-2 researcher under the program Produtividade em Pesquisa-CNPq, 307905/2014-4 and is a Procientista under SR2-UERJ. C.~P.~Felix is a PhD student supported by the program Ci{\^e}ncias sem Fronteiras - CNPq, 234112/2014-0.

\appendix
\section{Explicit evaluation of the spectral density}\label{appA}
In this Appendix, we compute at leading order the spectral density of the topological current correlator. Using
\begin{equation}
K_{\mu}=\frac{g^{2}}{16\pi^{2}}\varepsilon_{\mu\nu\lambda\sigma}A_{\nu}^{a}\left(\partial_{\lambda}A_{\sigma}^{a}+\frac{g}{2}f^{abc}A_{\lambda}^{b}A_{\sigma}^{c}\right),
\label{Chern_Simons_current}
\end{equation}
we obtain for the current-current correlator $\left<KK\right>$ at one loop order
\begin{equation}
 \braket{K_{\mu}(x)K_{\nu}(y)}=\frac{g^{4}}{16^{2}\pi^{4}}\varepsilon_{\mu\beta\rho\sigma}\varepsilon_{\nu\lambda\theta\alpha} \braket{A_{\beta}^{a}(x)\partial_{\rho}A_{\sigma}^{a}(x)A_{\lambda}^{d}(y)\partial_{\theta}A_{\alpha}^{d}(y)}.
 \label{KK_correlator}
\end{equation}
Now, in Fourier space
\begin{eqnarray}
  \braket{K_\mu(x) K_\nu(y)}&=&\frac{g^{4}}{16^{2}\pi^{4}}\varepsilon_{\mu\beta\rho\sigma}\varepsilon_{\nu\lambda\theta\alpha} \braket{A_{\beta}^{a}(x)\partial_{\rho}A_{\sigma}^{a}(x)A_{\lambda}^{d}(y)\partial_{\theta}A_{\alpha}^{d}(y)}\nonumber\\
  &=&\frac{g^{4}}{16^{2}\pi^{4}}\varepsilon_{\mu\beta\rho\sigma}\varepsilon_{\nu\lambda\theta\alpha}\int dkdpdudq \mathrm{e}^{iqy}\mathrm{e}^{ipx}\mathrm{e}^{ikx}\mathrm{e}^{iuy}p_{\rho}p_{\theta} \braket{A_{\beta}^{a}(k)A_{\sigma}^{a}(p)A_{\lambda}^{d}(u)A_{\alpha}^{d}(q)}.
\label{KK_cor_Fourier}
\end{eqnarray}
We perform the contractions using Wick's theorem and disregarding the disconnected contributions, we get
\begin{eqnarray}
 \braket{K_\mu(x) K_\nu(y)}=\frac{g^{4}(N^{2}-1)}{16^{2}\pi^{4}}\varepsilon_{\mu\beta\rho\sigma}\varepsilon_{\nu\lambda\theta\alpha}\left[\int dkdq \mathrm{e}^{i(x-y)(k-q)}q_{\rho}q_{\theta}D_{\beta\lambda}(k)D_{\sigma\alpha}(-q)+\right.\nonumber\\
 \left.+\int dkdp\mathrm{e}^{i(x-y)(p+k)}p_{\rho}k_{\theta}D_{\beta\alpha}(k)D_{\sigma\lambda}(p)\right].
 \label{KK_cor_Fourier_solution}
\end{eqnarray}
Via the substitutions
\begin{equation}
\ell=k-q \quad \mathrm{and} \quad \ell'=p+k,
\label{vai_chan}
\end{equation}
we can rewrite \eqref{KK_cor_Fourier_solution}
\begin{equation}
 \braket{K_\mu(x) K_\nu(y)}=\int\frac{d^{d}\ell}{(2\pi)^{d}}\mathrm{e}^{i\ell(x-y)}\mathcal{F}_{\mu\nu}(\ell)+\int\frac{d^{d}\ell'}{(2\pi)^{d}}\mathrm{e}^{i\ell'(x-y)}\mathcal{G}_{\mu\nu}(\ell'),
 \label{KK_cor_Fourier_transform}
\end{equation}
by which
\begin{equation}
\mathcal{F}_{\mu\nu}(\ell)=\frac{g^{4}(N^{2}-1)}{16^{2}\pi^{4}}\varepsilon_{\mu\beta\rho\sigma}\varepsilon_{\nu\lambda\theta\alpha}\int\frac{d^{d}k}{(2\pi)^{d}}\left[(k-\ell)_{\rho}(k-\ell)_{\theta}D_{\beta\lambda}(k)D_{\sigma\alpha}(\ell-k)\right]
\label{fl}
\end{equation}
and
\begin{equation}
\mathcal{G}_{\mu\nu}(\ell')=\frac{g^{4}(N^{2}-1)}{16^{2}\pi^{4}}\varepsilon_{\mu\beta\rho\sigma}\varepsilon_{\nu\lambda\theta\alpha}\int\frac{d^{d}k}{(2\pi)^{d}}\left[(\ell'-k)_{\rho}k_{\theta}D^{\beta\alpha}(k)D^{\sigma\lambda}(\ell'-k)\right].
\label{glprime}
\end{equation}
Using {\sc FeynCalc} \cite{feyncalc}, these expressions can be further simplified to
\begin{eqnarray}
\mathcal{F}_{\mu\nu}(\ell)=\frac{g^{4}(N^{2}-1)}{16^{2}\pi^{4}}\int\frac{d^{d}k}{(2\pi)^{d}}\frac{D(\ell-k)}{k^{2}}&&\left\{\delta_{\mu\nu}\left[k^{2}\left(\ell^{2}(D(k)-L(k))-2D(k)k^{2}\right)+\right.\right.\nonumber\\
&&+\left.\left.k.\ell\left((L(k)-D(k))k.\ell+4D(k)k^{2}\right)-2D(k)k^{2}\ell^{2}\right]+\right.\nonumber\\
&&+\left.
\ell_{\mu}\left[\vphantom{\frac{1}{2}}
\ell_{\nu}\left(k^{2}(L(k)-D(k))+2D(k)k^{2}\right)+\right.\right.\nonumber\\
&&+\left.\left.k_{\nu}\left((D(k)-L(k))k.\ell-2D(k)k^{2}\right)\vphantom{\frac{1}{2}}\right]+\right.\nonumber\\
&&+\left.k_{\mu}\left[\vphantom{\frac{1}{2}}k_{\nu}\left(\ell^{2}(L(k)-D(k))+2D(k)k^{2}\right)+\right.\right.\nonumber\\
&&+\left.\left.\ell_{\nu}\left((D(k)-L(k))k.\ell-2D(k)k^{2}\right)
\vphantom{\frac{1}{2}}\right]
\right\}
\label{flM}
\end{eqnarray}and
\begin{eqnarray}
\mathcal{G}_{\mu\nu}(\ell')=\frac{g^{4}(N^{2}-1)}{16^{2}\pi^{4}}\int\frac{d^{d}k}{(2\pi)^{d}}&&2D(k)D(\ell'-k)\left[\delta_{\mu\nu}(k.\ell'-k^{2})+k_{\mu}(k_{\nu}-\ell'_{\nu})\right].
\label{glprimeM}
\end{eqnarray}
We are only interested in the longitudinal component, which can be extracted by acting with the appropriate projector $L_{\mu\nu}(p)$. As it can be easily verified, the $\alpha$-dependent contributions do cancel, as expected from gauge invariance. As such, from here on, we can set $L(k)=0$. Doing so, \eqref{flM} reduces to
\begin{eqnarray}
\mathcal{F}_{\mu\nu}(\ell)=\frac{g^{4}(N^{2}-1)}{16^{2}\pi^{4}}\int\frac{d^{d}k}{(2\pi)^{d}}\frac{D(k)D(\ell-k)}{k^{2}}&&\left\{-\delta_{\mu\nu}\left[2k^{2}\ell^{2}+k^{2}(2k^{2}-\ell^{2})-4k^{2}k.\ell+(k.\ell)^{2}\right]+\right.\nonumber\\
&&+\left.\ell_{\mu}\left[k^{2}\ell_{\nu}+k_{\nu}(k.\ell-2k^{2})\right]+k_{\mu}\left[(2k^{2}-\ell^{2})k_{\nu}+\right.\right.\nonumber\\
&&+\left.\left.\ell_{\nu}(k.\ell-2k^{2})\right]
\right\}.
\label{flMLG}
\end{eqnarray}
Eventually, in Fourier space we obtain
\begin{eqnarray}
 \braket{K_\mu(p) K_\nu(-p)}&=&\frac{g^{4}(N^{2}-1)}{16^{2}\pi^{4}}\int\frac{d^{d}k}{(2\pi)^{d}}D(k)D(p-k) \Biggl\{6\delta_{\mu\nu}k.p+4k_{\mu}k_{\nu}-4\delta_{\mu\nu}k^{2}-\delta_{\mu\nu}p^{2}-\Biggr.\nonumber\\
&-&\Biggl.4k_{\mu}p_{\nu}+\delta_{\mu\nu}\frac{(k.p)^{2}}{k^{2}}+p_{\mu}p_{\nu}+p_{\mu}k_{\nu}\frac{k.p}{k^{2}}-2p_{\mu}k_{\nu}-p^{2}\frac{k_{\mu}k_{\nu}}{k^{2}}+k_{\mu}p_{\nu}\frac{k.p}{k^{2}}\Biggr\}
 \label{KK_correlator_simple}
\end{eqnarray}
The longitudinal part of \eqref{KK_correlator_simple} is then given by
\begin{eqnarray}
\braket{K_\mu(p) K_\nu(-p)}&=&\frac{g^{4}(N^{2}-1)}{16^{2}\pi^{4}}\frac{p_\mu p_\nu}{p^2}\int\frac{d^{d}k}{(2\pi)^{d}}D(k)D(p-k) 4\frac{(k.p)^{2}-k^{2}p^{2}}{p^{2}}\nonumber\\
&=&\frac{g^{4}(N^{2}-1)}{16^{2}\pi^{4}}\int\frac{d^{d}k}{(2\pi)^{d}}\frac{1}{k^{2}-m_{1}^{2}}\frac{1}{(p-k)^{2}-m_{2}^{2}}4\frac{(k.p)^{2}-k^{2}p^{2}}{p^{2}}.
\label{longitudinal_KK}
\end{eqnarray}
The next step is to rewrite \eqref{longitudinal_KK} in spectral form, i.e.~to extract the spectral density $\rho_\|(p^2)$, making use of
\begin{eqnarray}\label{decomp}
D(p^2)=\frac{p^2+M_1^2}{p^4+M_2^2p^2+M_3^4}=\frac{A_+}{p^2+m_+^2}+\frac{A_-}{p^2+m_-^2},
\end{eqnarray}
where we assume $M_2^4<4M_3^4$ as motivated from all lattice fits, so that the poles and residues are complex-conjugate numbers. Using this decomposition into standard massive Feynman propagators, each propagator with mass $m_+^2$ needs to be combined with an accompanying propagator with complex-conjugate mass $m_-^2$ to assure a branch cut along the usual (real) half-axis, consistent with a standard K\"all\'{e}n-Lehmann integral \cite{Baulieu:2009ha}.  This means that we need to take into account the following contributions,
\begin{eqnarray}
\braket{K_\mu(p) K_\nu(-p)}&=&\frac{g^{4}(N^{2}-1)}{16^{2}\pi^{4}}\int\frac{d^{d}k}{(2\pi)^{d}}\left(\frac{A_+}{k^{2}+m_{+}^{2}}\frac{A_-}{(p-k)^{2}+m_{-}^{2}}+\frac{A_-}{k^{2}+m_{-}^{2}}\frac{A_+}{(p-k)^{2}+m_{+}^{2}}\right)4\frac{(k.p)^{2}-k^{2}p^{2}}{p^{2}}\nonumber\\
&=&\frac{g^{4}(N^{2}-1)}{16^{2}\pi^{4}}\int\frac{d^{d}k}{(2\pi)^{d}}2A_+A_-\left(\frac{1}{k^{2}+m_{+}^{2}}\frac{1}{(p-k)^{2}+m_-^2}\right)4\frac{(k.p)^{2}-k^{2}p^{2}}{p^{2}}.
\label{longitudinal_KK10}
\end{eqnarray}
We will for the time being forget about the prefactor $2A_+A_-$ and will restore it at the end.

To continue,  we temporarily look  at a general massive propagator in Minkowski space, i.e. at
\begin{eqnarray}
\frac{g^{4}(N^{2}-1)}{16^{2}\pi^{4}}\int\frac{d^{d}k}{(2\pi)^{d}}\frac{1}{k^{2}-m_{1}^{2}}\frac{1}{(p-k)^{2}-m_{2}^{2}}4\frac{(k.p)^{2}-k^{2}p^{2}}{p^{2}}.
\label{longitudinal_KK2}
\end{eqnarray}
The reason is that, as discussed in \cite{Dudal:2010wn}, to compute the spectral density entering the K\"all\'{e}n-Lehmann representation, we can formally use the (Minkowski) Cutkosky cut rules \cite{Cutkosky:1960sp} pretending to work with particles with on-shell real masses, and at the end,  move back to Euclidean space, simultaneously replacing the masses with their respective complex-conjugate values  \begin{equation}
m_{1}^{2}\to m_+^2=a+ib \quad \mathrm{and} \quad m_{2}^{2}\to m_-^2=a-ib.
\label{cmp}
\end{equation}
So, let us apply the Cutkosky rules via the usual replacement,
\begin{equation}
\frac{1}{k^{2}-m_{1}^{2}}\rightarrow 2\pi\theta(k^{0})\delta({\vec{k}^{2}-m_{1}^{2}})
\label{p1_r}
\end{equation}
and
\begin{equation}
\frac{1}{(p-k)^{2}-m_{2}^{2}}\rightarrow 2\pi\theta((p-k)^{0})\delta({(\vec{p}-\vec{k})^{2}-m_{2}^{2}}).
\label{p2_r}
\end{equation}
We know that $\rho(\tau)\propto Disc \mathcal{K}(\tau)=2\text{Im}~\mathcal{K}(\tau)$, thus
\begin{eqnarray}
\text{Im}~\mathcal{K}(p,m_{1},m_{2})&=&\frac{1}{2}\frac{g^{4}(N^{2}-1)}{16^{2}\pi^{4}}\int\frac{d^{d}k}{(2\pi)^{d-2}}\theta(k^{0})\delta({\vec{k}^{2}-m_{1}^{2}})\theta((p-k)^{0})\delta({(\vec{p}-\vec{k})^{2}-m_{2}^{2}})\left(\frac{4(k.p)^{2}}{p^{2}}-\right.\nonumber\\
&&\left.-4k^{2}\vphantom{\frac{1}{2}}\right).
\label{ImF}
\end{eqnarray}
We will work in the center-of-mass frame, i.e.~$p_{\mu}=(p^{0},0)=(E,0)$, then
\begin{eqnarray}
\text{Im}~\mathcal{K}(E^{2})&=&\frac{1}{2}\frac{g^{4}(N^{2}-1)}{16^{2}\pi^{4}}\int\frac{d^{d}k}{(2\pi)^{d-2}}\theta(k^{0})\delta({(\vec{k}^{0})^{2}-\omega_{k,1}^{2}})\theta((E-k)^{0})\delta({(E-k^{0})^{2}-\omega_{k,2}^{2}})\left(\frac{4(k^{0}E)^{2}}{E^{2}}-\right.\nonumber\\
&&\left.-4((k^{0})^{2}+\vec{k}^{2})\vphantom{\frac{1}{2}}\right)\nonumber\\
&=&\frac{1}{2}\frac{g^{4}(N^{2}-1)}{16^{2}\pi^{4}}\int\frac{d^{d}k}{(2\pi)^{d-2}}\theta(k^{0})\delta({(\vec{k}^{0})^{2}-\omega_{k,1}^{2}})\theta((E-k)^{0})\delta({(E-k^{0})^{2}-\omega_{k,2}^{2}})\left(4(k^{0})^{2}-\right.\nonumber\\
&&\left.-4(k^{0})^{2}-4\vec{k}^{2}\right)\nonumber\\
&=&-2\frac{g^{4}(N^{2}-1)}{16^{2}\pi^{4}}\int\frac{d^{d}k}{(2\pi)^{d-2}}\theta(k^{0})\delta({(\vec{k}^{0})^{2}-\omega_{k,1}^{2}})\theta((E-k)^{0})\delta({(E-k^{0})^{2}-\omega_{k,2}^{2}})\vec{k}^{2},
\label{ImF_CoM}
\end{eqnarray}
with $\omega_{k,i}=\sqrt{\vec{k}^{2}+m_{i}^{2}}$.
Integrating over $k^{0}$ and changing for spherical coordinates, we obtain
\begin{eqnarray}
\text{Im}~\mathcal{K}(E^{2})=-\frac{\pi^{d-1/2}}{\Gamma(\frac{d-1}{2})}\frac{g^{4}(N^{2}-1)}{16^{2}\pi^{4}}\int\frac{d|\vec{k}|}{(2\pi)^{d-2}}|\vec{k}|^{d-2}\frac{1}{\omega_{k,1}\omega_{k,2}}\delta(E-\omega_{k,1}-\omega_{k,2})\vec{k}^{2}.
\label{ImF_k_0}
\end{eqnarray}
Using the property $\delta(g(\vec{k}^{2}))=\frac{1}{|g\prime(\vec{k}_{0}^{2})|}\delta(\vec{k}^{2}-\vec{k}_{0}^{2})=\frac{1}{2|\vec{k}_{0}||g\prime(\vec{k}_{0}^{2})|}\delta(|\vec{k}|-|\vec{k}_{0}|)$, where $\vec{k}_{0}^{2}$ is such that $g(\vec{k}_{0}^{2})=0$, i.e.
\begin{eqnarray}
g(\vec{k}_{0}^{2})&=&E-\sqrt{\vec{k}_{0}^{2}+m_{1}^{2}}-\sqrt{\vec{k}_{0}^{2}+m_{2}^{2}}=0\nonumber\\
&|\vec{k}_{0}|&=\frac{\sqrt{(E^{2}-m_{1}^{2}-m_{2}^{2})^{2}-4m_{1}^{2}m_{2}^{2}}}{2E},
\label{k_0}
\end{eqnarray}
\eqref{ImF_k_0} gives us
\begin{eqnarray}
\text{Im}~\mathcal{K}(E^{2})=-\frac{g^{4}(N^{2}-1)}{2^{d+6}\pi^{5/2}\Gamma(\frac{d-1}{2})}\frac{|\vec{k}_{0}|^{d-3}\vec{k}_{0}^{2}}{E}.
\label{ImF_solved}
\end{eqnarray}
With conjugate masses parameterized as in \eqref{cmp}, \eqref{k_0} can be rewritten as
 \begin{equation}
 |\vec{k}_{0}|=\frac{\sqrt{E^{4}-4b^{2}-4aE^{2}}}{2E}
 \label{k_0_cmp}
 \end{equation}
and with the two last equations, we can write \eqref{ImF_solved} as
\begin{eqnarray}
\text{Im}~\mathcal{K}(E^{2})&=&-\frac{g^{4}(N^{2}-1)}{2^{d+6}\pi^{5/2}\Gamma(\frac{d-1}{2})}\frac{\left(\frac{\sqrt{E^{4}-4b^{2}-4aE^{2}}}{2E}\right)^{d-3}\left(\frac{\sqrt{E^{4}-4b^{2}-4aE^{2}}}{2E}\right)^{2}}{E}\nonumber\\
&=&-\frac{g^{4}(N^{2}-1)}{2^{2d+5}\pi^{5/2}\Gamma(\frac{d-1}{2})}\frac{\left(E^{4}-4b^{2}-4aE^{2}\right)^{(d-1)/2}}{E^{d}}.
\label{ImF_solved_cmp}
\end{eqnarray}
Using $E^{2}\rightarrow \tau$ and the equivalence $\rho=\frac{1}{\pi}\text{Im}~ \mathcal{K}(\tau)$ we finally get
\begin{eqnarray}
\rho_\|(\tau)=-2A_+A_-\frac{g^{4}(N^{2}-1)}{2^{2d+5}\pi^{7/2}\Gamma(\frac{d-1}{2})}\frac{\left(\tau^{2}-4b^{2}-4a\tau\right)^{(d-1)/2}}{\tau^{d/2}}\qquad\text{for}~\tau\geq \tau_c.
\label{rho_tau}
\end{eqnarray}
The threshold is given by \cite{Dudal:2010wn}
\begin{eqnarray}\label{threshold}
\tau_c=(m_++m_-)^2=2(a+\sqrt{a^2+b^2}).
\end{eqnarray}
We also restored the prefactor $2A_+A_-$.

\end{document}